\documentclass[referee]{raa}            

\usepackage{graphicx,times}             
\usepackage{natbib}
\usepackage{amssymb,amsmath}
\bibpunct{(}{)}{;}{a}{}{,}

\usepackage[a4paper=true,pagebackref=true,bookmarks=false]{hyperref}
\hypersetup{colorlinks = true, linkcolor = green, anchorcolor = red, citecolor = blue, filecolor = red, pagecolor = red, urlcolor = red}

\def\jref@jnl#1{{\rm#1\/}}
\def\actaa{\jref@jnl{Acta Astronomica}}
\def\aap{\jref@jnl{A\&A}}
\def\aapr{\jref@jnl{The Astronomy and Astrophysics Review}}
\def\aaps{\jref@jnl{Astronomy and Astrophysics Supplement Series}}
\def\aj{\jref@jnl{AJ}}
\def\apj{\jref@jnl{ApJ}}
\def\apjl{\jref@jnl{ApJL}}
\def\apjs{\jref@jnl{ApJS}}
\def\apss{\jref@jnl{Astrophysics and Space Science}}
\def\ao{\jref@jnl{Applied Optics}}
\def\araa{\jref@jnl{ARA\&A}}
\def\bain{\jref@jnl{BAN}}
\def\caa{\jref@jnl{Chinese Astronomy and Astrophysics}}
\def\cjaa{\jref@jnl{Chinese Journal of Astronomy and Astrophysics}}
\def\gca{\jref@jnl{Geochimica et Cosmochimica Acta}}
\def\jcp{\jref@jnl{Journal of Chemical Physics}}
\def\jqsrt{\jref@jnl{Journal of Quantitative Spectroscopy and Radiative Transfer}}
\def\mnras{\jref@jnl{MNRAS}}
\def\memras{\jref@jnl{Memoirs of the Royal Astronomical Society}}
\def\memsai{\jref@jnl{Memorie della Societa Astronomica Italiana}}
\def\na{\jref@jnl{New Astronomy}}
\def\nar{\jref@jnl{New Astronomy Reviews}}
\def\nat{\jref@jnl{Nature}}
\def\pasa{\jref@jnl{Publications of the Astronomical Society of Australia}}
\def\planss{\jref@jnl{Planetary and Space Science}}
\def\pasj{\jref@jnl{Publications of the Astronomical Society of Japan}}
\def\pasp{\jref@jnl{PASP}}
\def\physrep{\jref@jnl{Physics Reports}}
\def\pra{\jref@jnl{Physical Review A}}
\def\prd{\jref@jnl{Physical Review D}}
\def\pre{\jref@jnl{Physical Review E}}
\def\physrep{\jref@jnl{Physics Reports}}
\def\physscr{\jref@jnl{Physica Scripta}}
\def\qjras{\jref@jnl{Quarterly Journal of the Royal Astronomical Society}}
\def\rmxaa{\jref@jnl{Revista Mexicana de Astronomia y Astrofisica}}
\def\skytel{\jref@jnl{Sky and Telescope}}
\def\solphys{\jref@jnl{Solar Physics}}
\def\sovast{\jref@jnl{Soviet Astronomy}}
\def\ssr{\jref@jnl{Space Science Reviews}}
\def\zap{\jref@jnl{Zeitschrift fuer Astrophysik}}
\def\azh{\jref@jnl{Astronomicheskij Zhurnal}}
\def\procspie{\jref@jnl{Proc. SPIE}}

\def\akn{}

\newcommand{\MC}{\multicolumn}

\begin{document}

   \title{Searching For Wide Binary Stars with Non-coeval Components in the Southern Sky}

   \volnopage{Vol.0 (20xx) No.0, 000--000}      
   \setcounter{page}{1}          

   \author{Alexey Kniazev
      \inst{1,2,3,4}
   \and Oleg Malkov
      \inst{5}
   }

   \institute{South African Astronomical Observatory, Cape Town, 7935, South Africa {\it a.kniazev@saao.nrf.ac.za}\\
        \and
              Southern African Large Telescope, Cape Town, 7935, South Africa\\
        \and
              Special Astrophysical Observatory, Nizhnij Arkhyz, Karachai-Circassia, 369167, Russia\\
        \and 
              Sternberg State Astronomical Institute, Moscow, Universitetsky ave., 13, Russia\\
        \and
              Institute of Astronomy, Russian Academy of Sciences, 48 Pyatnitskaya St., Moscow 119017, Russia\\
\vs\no
   {\small Received~~20xx month day; accepted~~20xx~~month day}}

\abstract{We have completed our observational program to search for wide binary systems
 with non-coeval components in the southern sky and report our results here.
 The final set of four systems was spectroscopically investigated in this paper. No binary systems
 with components of different ages were found among them. Taking into account our previous studies,
 we estimate the fraction of such binaries (i.e., binaries formed, presumably, by capture)
 not higher than 0.06~\%.
 The study will be continued on the northern sky.
}

   \authorrunning{A. Yu. Kniazev \& O. Yu. Malkov}  
   \titlerunning{Wide binary stars with non-coeval components}  

   \maketitle


%
%

\section{Introduction}

According to current understanding, the formation of binary stars follows one of two scenarios: the fission of rotating clouds of molecular gas during the gravitational collapse and inelastic collisions of stars during the formation of young star clusters~\citep{2020PhyU...63..209T}. A possible (but obviously rare) phenomenon could be the formation of a binary star by capture. Capture occurs when two stars pass close to each other. There must necessarily be a scattering medium (e.g. a circumstellar disk or a third star) to which the excess kinetic energy can be transferred. 

Obviously, a sufficient (but not necessary) observational condition to demonstrate a binary system was formed by capture is the difference in age of the components. The age of a star is a rather difficult parameter to define. However, the presence in a wide binary system (where the transfer of matter between the components is excluded) of a more evolved and yet less massive component is irrefutable evidence for an age difference in the components. In our previous attempt to find such pairs, we compared the lifetime of a star on the main sequence (MS) and on the pre-MS stage and found three candidates~\citep{2000IAUS..200P.170M}.

In this study, we compare two-dimensional spectral classes and component masses estimated from the spectral classification. The situation where the less massive component evolved further from the MS can be considered as a binary system formed by capture.

The structure of this article is as follows: Section~\ref{txt:sample} describes our sample selection.
Data from the literature on the systems under study are presented in Section~\ref{txt:general_info}.
Section~\ref{txt:spec_obs} describes our observations and reduction of the spectral data.
Data analysis is described in Section~\ref{txt:analysis} and the results are discussed in Section~\ref{txt:results}. Section~\ref{txt:summary} summarizes this article.

\section{Sample selection}
\label{txt:sample}

Comparing the spectral classes and component masses (estimated from the spectral classification) of binary systems from the Sixth Catalog of Orbits of Visual Binary Stars, ORB6~\citep{2001AJ....122.3472H}, we found thirteen systems in which the less massive component appears more evolved and, therefore, the pair was probably formed by capture~\citep{2020INASR...5..341M}.

For these systems, data available in the literature and other catalogues and databases were further studied: Washington Double Star Catalog \citep[WDS;][]{2001AJ....122.3466M}, Catalog of Stellar Spectral Classifications~\citep{2014yCat....102023S}, Multiple star catalog, \citep[MSC;][]{2018ApJS..235....6T}, Binary star database, \citep[BDB;][]{2015A&C....11..119K} as well as in the SIMBAD database.
Our spectral observations of three of these systems and their analysis revealed that at least one of the systems, HD~156331, could have formed by capture~\citep{2022OAst...31..327M}. Four other systems out of those listed in \citet{2020INASR...5..341M} are the subject of the present study. Their parameters are listed in Table~\ref{tab:pairs}.

\begin{table}
    \akn
    \caption{Parameters of the systems under study and observational log}
    \label{tab:pairs}
    \begin{center}
        \begin{tabular}{l|r|r|c|c|c|r}
            \hline
            Name       &   $V$    &\MC{1}{c|}{$\varpi$} &   Date        &   Exposure     &   Seeing   &   SNR      \\     
                       &   (mag)  &\MC{1}{c|}{(mas)}    &               &\MC{1}{c|}{(s)} &   (arcsec) &            \\
            \hline                                                                                             
            HD\,~20121 &  5.93    &     23.53$\pm$0.62  &  2021.07.24   & 3$\times$40    &    2.0     &  160        \\     
            HD\,~26722 &  5.05    &      9.83$\pm$0.64  &  2021.09.04   & 3$\times$25    &    1.5     &  200       \\     
            HD\,114330 &  4.40    &     11.18$\pm$0.41  &  2021.07.11   & 3$\times$15    &    1.8     &  240       \\     
            HD\,187949 &  6.48    &      8.67$\pm$0.14  &  2021.07.10   & 3$\times$45    &    1.4     &  190       \\     
            \hline                       
        \end{tabular}
    \end{center}
Visual brightness $V$ is taken
from The Bright Star Catalogue~\citep{1991bsc..book.....H}
for HD\,26722,
and from SIMBAD for HD\,20121, HD\,114330, HD\,187949.
Parallax $\varpi$ is referenced from Hipparcos~\citep{2007A&A...474..653V} for HD\,20121, HD\,26722,
and from Gaia DR3~\citep{2016A&A...595A...1G, 2022arXiv220800211G} for HD\,114330, HD\,187949.
\end{table}

\section{Known Information About Studied Systems}
\label{txt:general_info}

\subsection{HD\,20121 = WDS\,03124-4425}
\label{txt:20121}

HD\,20121\,AB ($6^m.50+6^m.90$, F7III+A0V) has the orbital
solution with semi-major axis and orbital period of a~=~0."410 and P~=~45.2~yr respectively.
It has a remote $8^m.91$ companion at separation $\rho$=3.8"
(Gaia DR3 parallax $\varpi$=22.74$\pm$0.02 mas) with common proper motion.
Apparently, its presence does not affect our spectral observations 
(the diameter of the fiber entrance was 2.23~arcsec during our spectral observations
as described in Section~\ref{txt:spec_obs}).

\subsection{HD\,26722 = WDS\,04139+0916 = 47\,Tau}
\label{txt:26722}

HD\,26722 ($5^m.05+7^m.32$, G5IV+A8V) has the orbital
solution with the semi-major axis and orbital period of a~=~1."053 and P~=~479~yr respectively.
No evidence of additional components in this binary system was found in the literature.
Note that the Gaia DR3 parallax $\varpi$=7.62$\pm$0.11 mas differs from the Hipparcos one
of $\varpi$=9.83$\pm$0.64, which is included in SIMBAD and listed in Table~\ref{tab:pairs}.

\subsection{HD\,114330 = WDS\,13099-0532 = $\theta$\,Vir}
\label{txt:114330}

HD\,114330 is a spectroscopic (S1) binary ($4^m.50+6^m.84$, A1IVs+Am).
Various orbital solutions for this pair were published in the literature.
According to comments, received from
the author of the MSC~\citep{2018ApJS..235....6T},
Dr.~A.~Tokovinin, the solution with the semi-major axis and orbital period of
a~=~0."219 and P~=~33~yr is certainly wrong, and the solution
a~=~0."116 and P~=~17.8~yr~\citep{1977RMxAA...3..109B} is inconsistent
with speckle observations (Tokovinin 2022, private communication).

HD\,114330 has a remote companion at $\rho$=7" (V=8~mag) with common proper motion
and orbital solution with a~=~1."243 and P~=~695~yr~\citep{2015IAUDS.185....1Z},
though Tokovinin (2022, private communication) considers this solution to be unreliable.
There is also another remote companion at $\rho$=71" (V=10.38 mag, K2III),
which appears to be an optical companion with HD\,114330).
Apparently, the presence of these two remote companions
does not affect our spectral observations.

\subsection{HD\,187949 = WDS\,19531-1436 (HD\,187949\,A = V505\,Sgr)}
\label{txt:187949}

HD\,187949\,AB is a wide pair ($6^m.58+9^m.13$, A2V+F7V)
with orbital period P=349.2 yr and semi-major axis a=0.68".
A brighter component, HD\,187949 A, is a close (spectroscopic S2 + eclipsing)
binary ($6^m.72+8^m.88$, A2V+G8IV, P=1.183~d) with an eclipse timing variation
with P=38.4~yr, caused probably by another, unseen companion at a~=~0.157".

\section{Observations and Data Reduction}
\label{txt:spec_obs}

All four systems from this paper were observed with
the High Resolution Spectrograph
\citep[HRS;][]{2008SPIE.7014E..0KB,2010SPIE.7735E..4FB,2012SPIE.8446E..0AB,2014SPIE.9147E..6TC},
which is a thermostabilized double-beam \'echelle spectrograph
installed at the
Southern African Large Telescope \citep[SALT;][]{2006SPIE.6267E..0ZB,2006MNRAS.372..151O}.
The acquired spectrum covered the total spectral range 3735--8870~\AA\ that
consist of the blue arm spectrum (3735--5580~\AA) and the red arm spectrum (5415--8870~\AA).
The spectrograph can be used in the low (LR, R$\approx$14,000-15,000)
medium (MR, R$\approx$40,000--43,000) and high (HR, R$\approx$67,000--74,000) resolution modes.
It is equipped with two fibers (object and sky fibers) for each mode.
For our observation, we used HRS in MR mode, where both the object and sky
fibers are 2.23~arcsec in diameter.
All additional details of observations are summarized in the Table~\ref{tab:pairs}.
Generally, each star was observed during one night with three exposures.
Exposures were selected in a way to accumulate Signal-to-Noise Ratio (SNR) of more than 150
in the spectral region 4300--8800~\AA. Unfortunately, the sensitivity of HRS drops down fast bluer of
4300~\AA\ and the final SNR in this spectral region is very hard to predict.

The HRS calibration plan consists of three spectral flats and one spectrum of the ThAr lamp
that were obtained in each mode weekly, which is enough to get average external accuracy of about
300 m~s$^{-1}$.
The method of analysis, described in Section~\ref{txt:analysis} needs
to use spectra corrected for the sensitivity curve.
For this reason spectra of spectrophotometric standards from the list
of \citet{2017SALT.REPORT7}\footnote{\url{https://astronomers.salt.ac.za/software/hrs-pipeline/}}
were observed and used during HRS data reduction.

Primary reduction of the HRS data, including overscan correction,
bias subtractions and gain correction, was done with the SALT
science pipeline \citep{2010SPIE.7737E..25C}.
Spectroscopic reduction of the HRS data was carried out using
standard HRS pipeline and our own additions to it,
as described in detail in \citep{2019AstBu..74..208K}.

\section{Spectral Data Analysis}
\label{txt:analysis}

\begin{table*}
    \akn
    \caption{Stellar parameters found with {\tt fbs} software}
    \label{tab:stars}
     \begin{tabular}{lrcrrcrrcl} \hline
System        & T$_{\rm eff}$ & $\log g$      & $\rm V \sin i$  & V$_{\rm hel}$  & Weight in        & $M_V$           & $\rm [Fe/H]$     & $\rm E(B-V)$  & St. lib.    \\
              &   (K)         & (cm~s$^{-2}$) & (km~s$^{-1}$)   & (km~s$^{-1}$)  & V band           & (mag)           &   (dex)          &   (mag)       &             \\
\hline                                                                                                                                                              
HD\,20121  A  & 6692$\pm$20   & 4.36$\pm$0.05  &  77.2$\pm$0.9  & $-$7.8$\pm$0.5 & 0.60$\pm$0.09    &  3.34$\pm$0.11  & $-$0.22$\pm$0.05 & 0.00$\pm$0.01 & Coelho      \\
HD\,20121  A  & 6540$\pm$10   & 4.14$\pm$0.03  &  79.6$\pm$2.4  &$-$10.4$\pm$0.5 & 0.51$\pm$0.01    &  3.52$\pm$0.11  & $-$0.22$\pm$0.02 & 0.00$\pm$0.01 & Phoenix     \\
\hline                                                                                                          
HD\,20121  B  & 6500$\pm$15   & 4.35$\pm$0.16  &  89.5$\pm$0.8  &   56.3$\pm$3.8 & 0.40$\pm$0.09    &  3.78$\pm$0.11  & $-$0.22$\pm$0.05 & 0.00$\pm$0.01 & Coelho      \\
HD\,20121  B  & 6473$\pm$10   & 4.15$\pm$0.04  &  76.2$\pm$1.5  &   38.6$\pm$2.6 & 0.49$\pm$0.01    &  3.56$\pm$0.11  & $-$0.22$\pm$0.02 & 0.00$\pm$0.01 & Phoenix     \\
\hline\hline                                                                                                                                                    
HD\,26722  A  & 7980$\pm$10   & 3.50$\pm$0.01  &  29.5$\pm$0.5  &   27.8$\pm$1.0 & 0.10$\pm$0.01    &  2.30$\pm$0.28  & $-$0.36$\pm$0.01 & 0.00$\pm$0.01 & Coelho      \\
HD\,26722  A  & 7600$\pm$25   & 3.10$\pm$0.08  &  26.1$\pm$0.7  &    2.4$\pm$0.4 & 0.16$\pm$0.01    &  1.79$\pm$0.22  & $-$0.33$\pm$0.01 & 0.00$\pm$0.01 & Phoenix     \\
\hline                                                                                                          
HD\,26722  B  & 5000$\pm$10   & 2.04$\pm$0.01  &   1.3$\pm$0.2  & $-$8.0$\pm$0.1 & 0.90$\pm$0.01    & -0.83$\pm$0.28  & $-$0.36$\pm$0.01 & 0.00$\pm$0.01 & Coelho      \\
HD\,26722  B  & 5100$\pm$10   & 2.06$\pm$0.01  &   5.6$\pm$0.2  & $-$8.0$\pm$0.2 & 0.84$\pm$0.01    & -0.01$\pm$0.22  & $-$0.33$\pm$0.01 & 0.00$\pm$0.01 & Phoenix     \\
\hline\hline                                                                                                                                                     
HD\,114330 A  & 9550$\pm$10   & 3.62$\pm$0.01  &  0.0$\pm$0.1   &$-$1.5$\pm$0.3  & 1.00$\pm$0.01    & -0.36$\pm$0.09  &    0.04$\pm$0.01 & 0.00$\pm$0.01 & Coelho      \\
HD\,114330 A  & 9350$\pm$10   & 3.54$\pm$0.01  &  2.8$\pm$0.1   &$-$1.6$\pm$0.3  & 1.00$\pm$0.01    & -0.36$\pm$0.09  &    0.18$\pm$0.05 & 0.00$\pm$0.01 & Phoenix     \\
\hline\hline                                                                                                          
HD\,187949 A  & 8920$\pm$30   & 4.36$\pm$0.03  & 91.8$\pm$0.5   &$-$41.0$\pm$0.6 & 0.90$\pm$0.01    &  1.28$\pm$0.25  &$-$0.29$\pm$0.01  & 0.00$\pm$0.01 & Coelho      \\ 
HD\,187949 A  & 8380$\pm$10   & 4.32$\pm$0.01  & 89.7$\pm$0.8   &$-$38.8$\pm$0.7 & 0.91$\pm$0.01    &  1.27$\pm$0.27  &$-$0.31$\pm$0.02  & 0.00$\pm$0.01 & Phoenix     \\ 
\hline                                                                                                                                                              
HD\,187949 B  & 6130$\pm$25   & 3.72$\pm$0.04  &  0.6$\pm$0.3   &  1.7$\pm$0.3   & 0.10$\pm$0.01    &  3.67$\pm$0.25  &$-$0.29$\pm$0.01  & 0.00$\pm$0.01 & Coelho      \\
HD\,187949 B  & 6280$\pm$10   & 3.52$\pm$0.13  &  6.3$\pm$0.6   &  1.9$\pm$0.1   & 0.09$\pm$0.01    &  3.78$\pm$0.27  &$-$0.31$\pm$0.02  & 0.00$\pm$0.01 & Phoenix     \\
\hline\hline                                                                                                                                                    
  Errors     &   300.         & 0.15          &  5.             &    1.3         &       0.02       &    ---          &   0.05            &        0.01   &            \\
\hline
    \end{tabular}
\end{table*}

\begin{figure*}[]
    \includegraphics[width=0.5\textwidth]{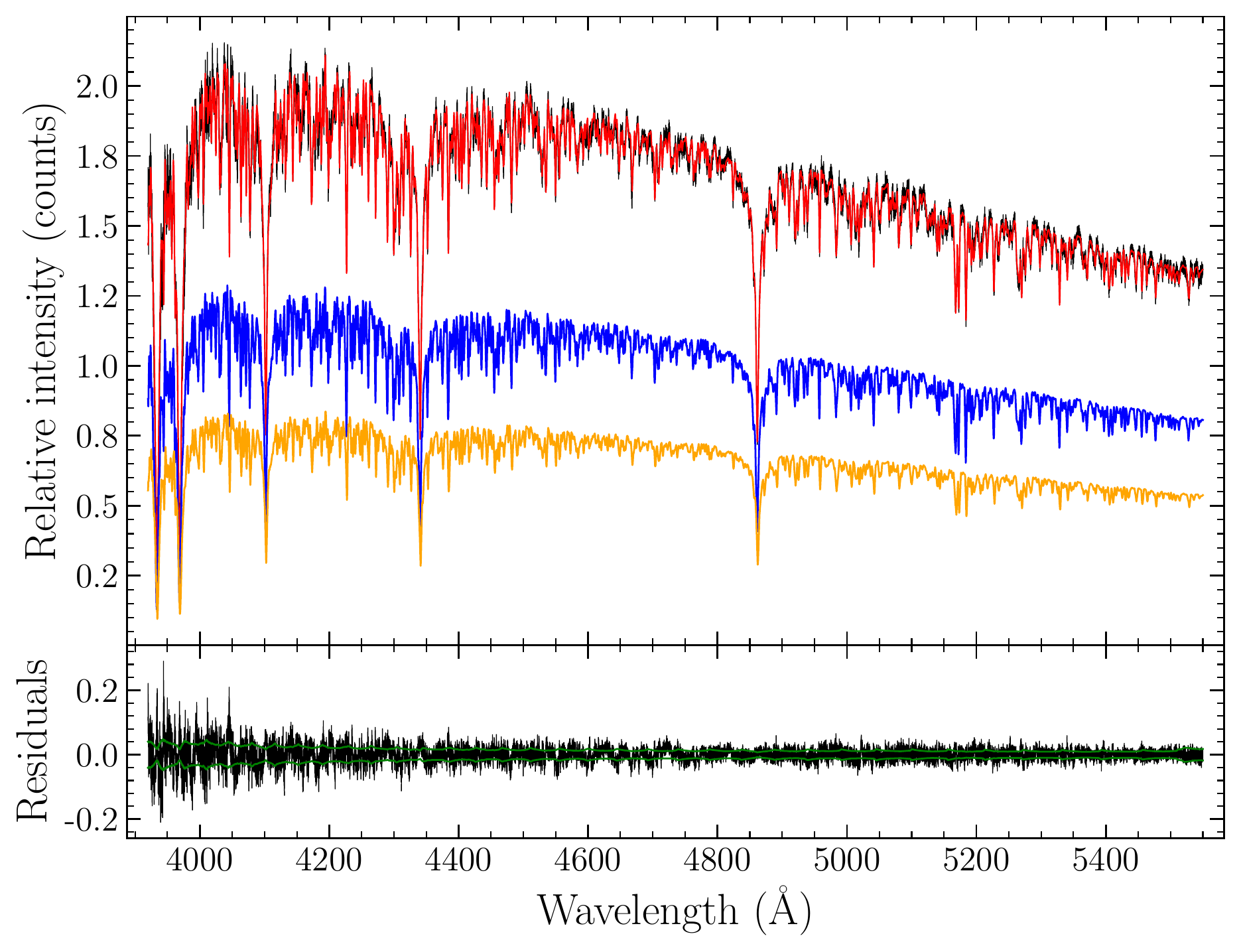}
    \includegraphics[width=0.5\textwidth]{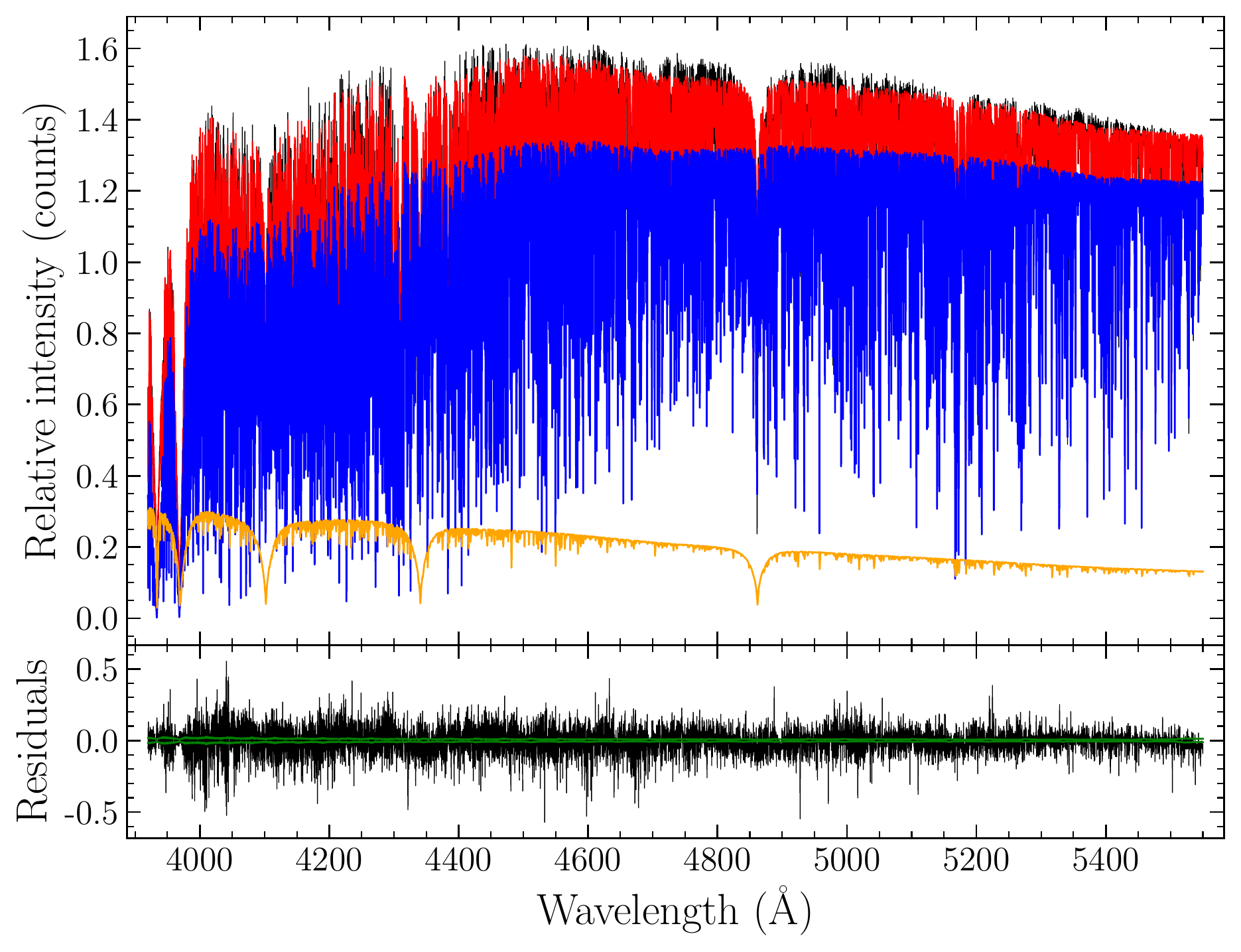}
    \includegraphics[width=0.5\textwidth]{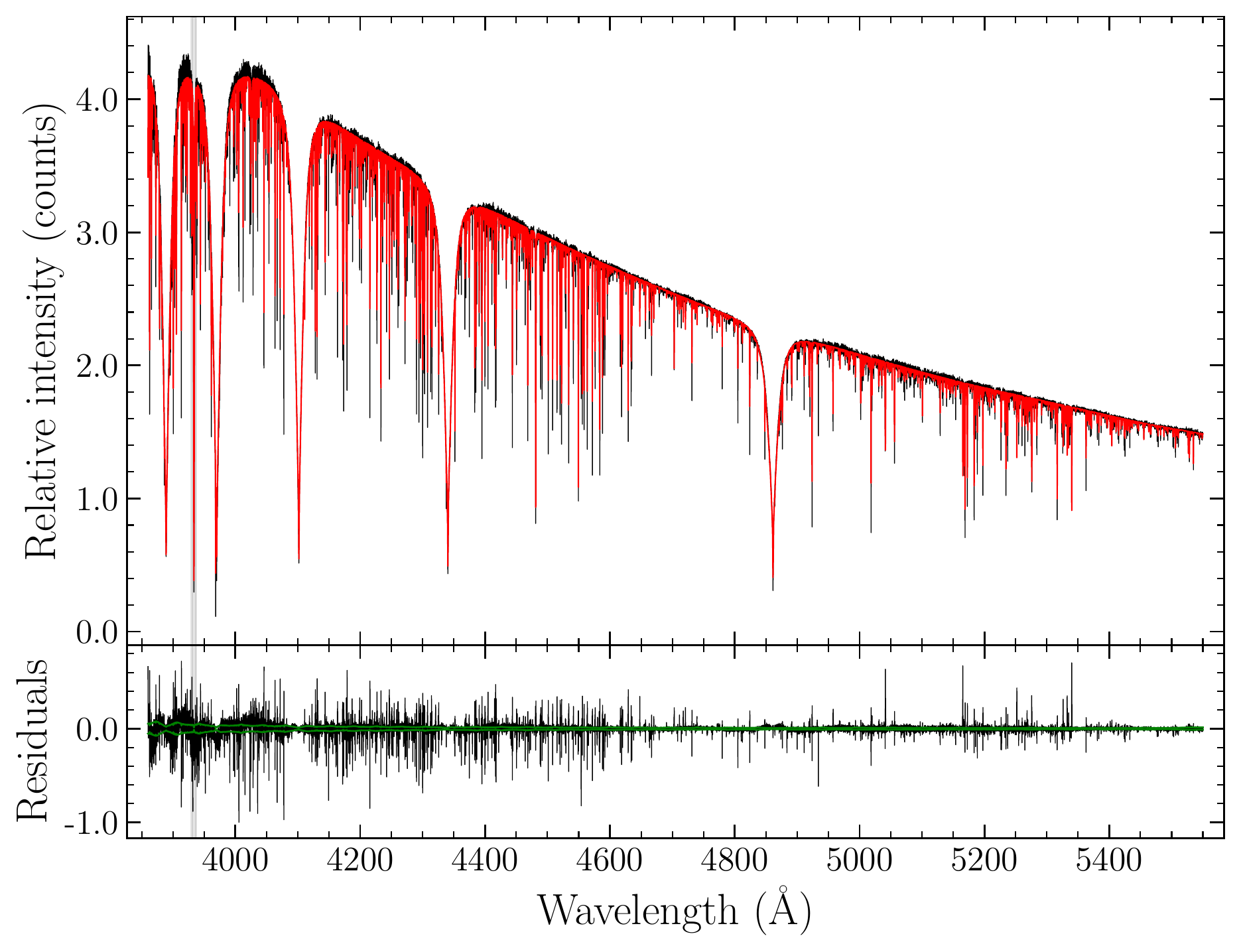}
    \includegraphics[width=0.5\textwidth]{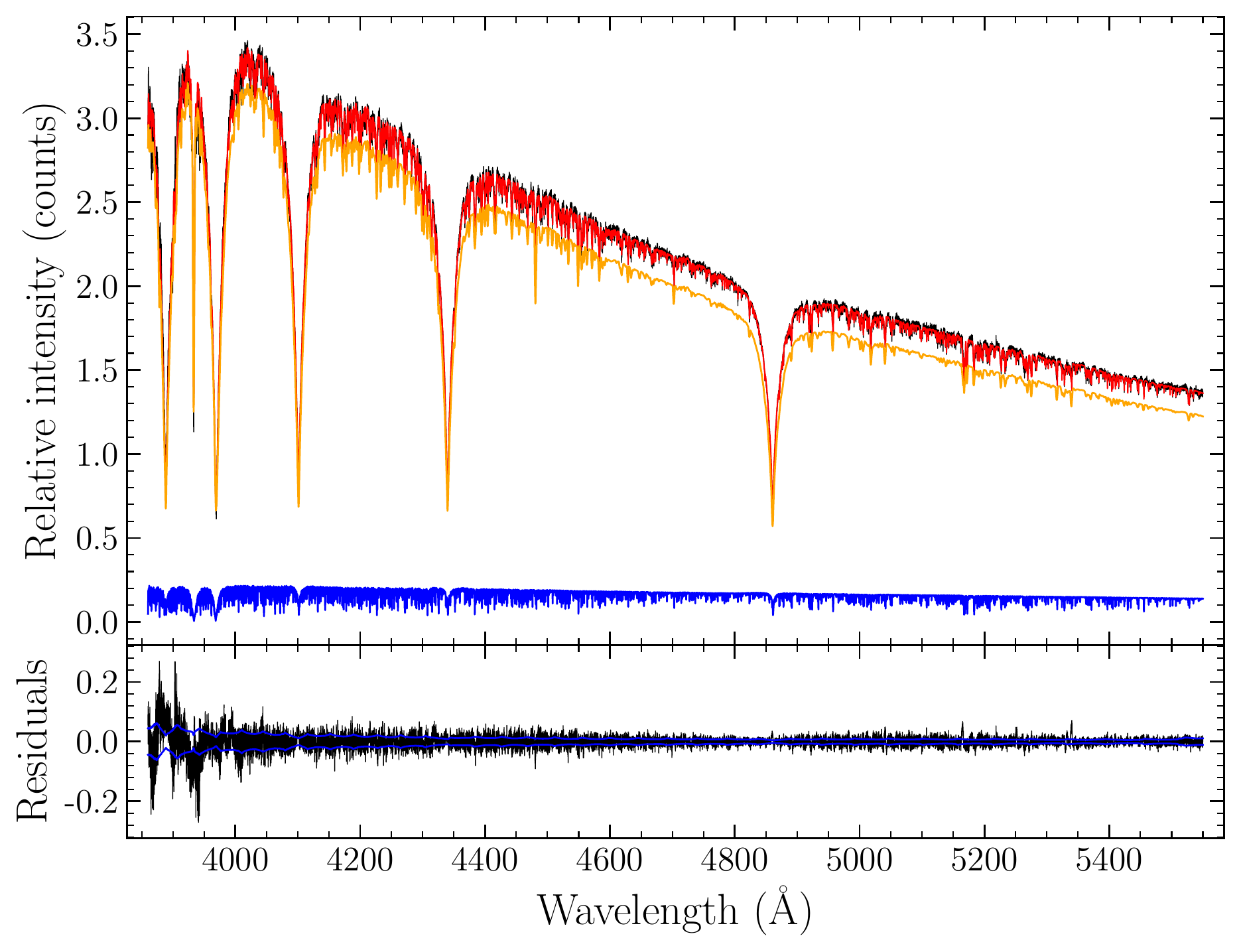}
    \caption{\akn The results of the fit all stars from this work by {\tt fbs} software:
HD\,20121    (top left),
HD\,26722    (top right)
HD\,114330   (bottom left),
HD\,187949   (bottom right).
             Each panel consists of two sub-panels: 
        {\bf the top one} features the result of the fit in the spectral region 3900-5550~\AA.
        The observed spectrum is drawn in black.
        Both found components are shown in blue and orange respectively, 
        whereas their sum is depicted in red;
        {\bf the bottom one} represents the difference between the observed spectrum and 
        its model in black,
        together with errors that were propagated from the HRS data reduction (continuous green line).
    }
    \label{fig:spectrum}
\end{figure*}

\begin{figure}[]
    \includegraphics[width=0.5\textwidth]{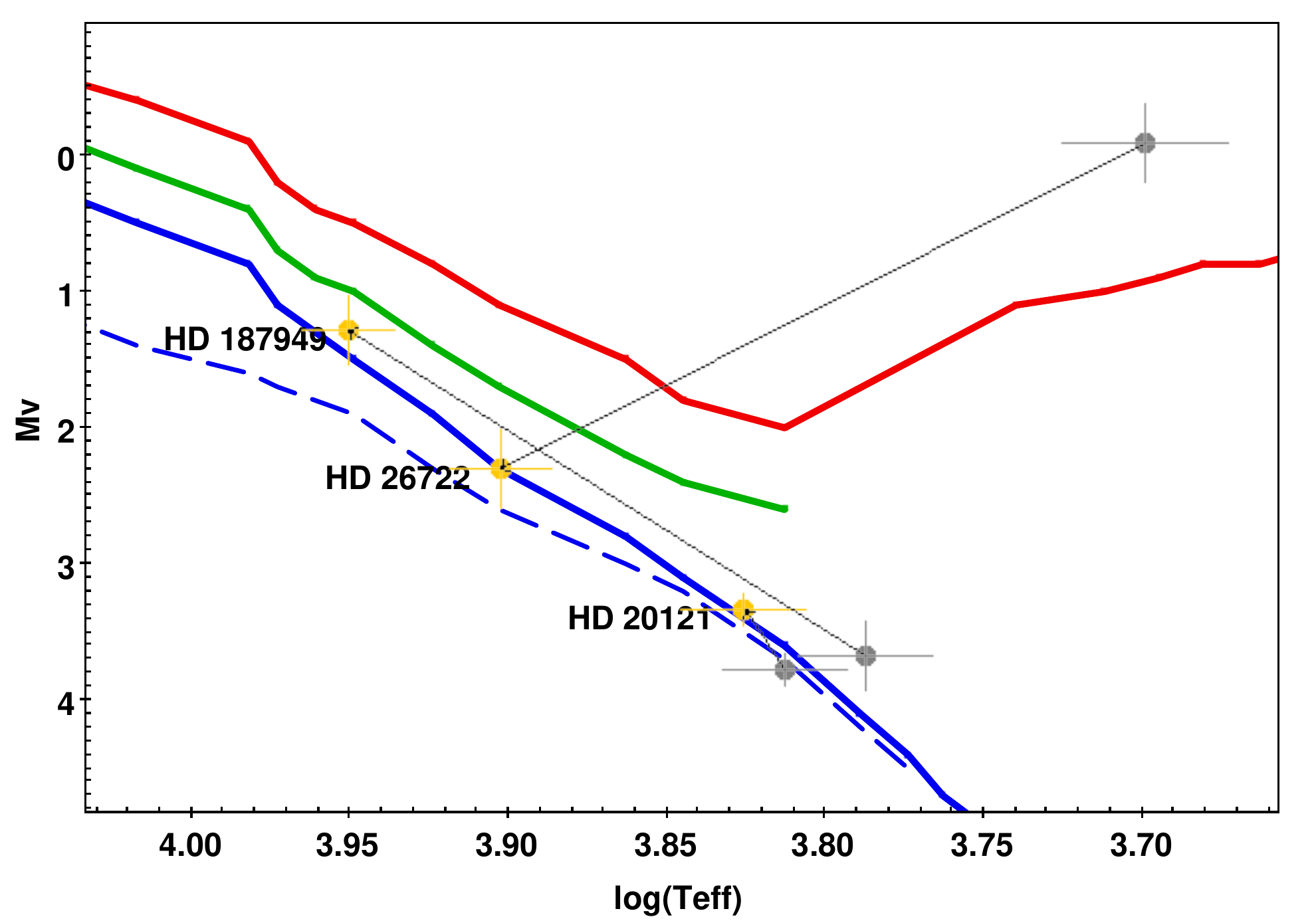}
    \includegraphics[width=0.5\textwidth]{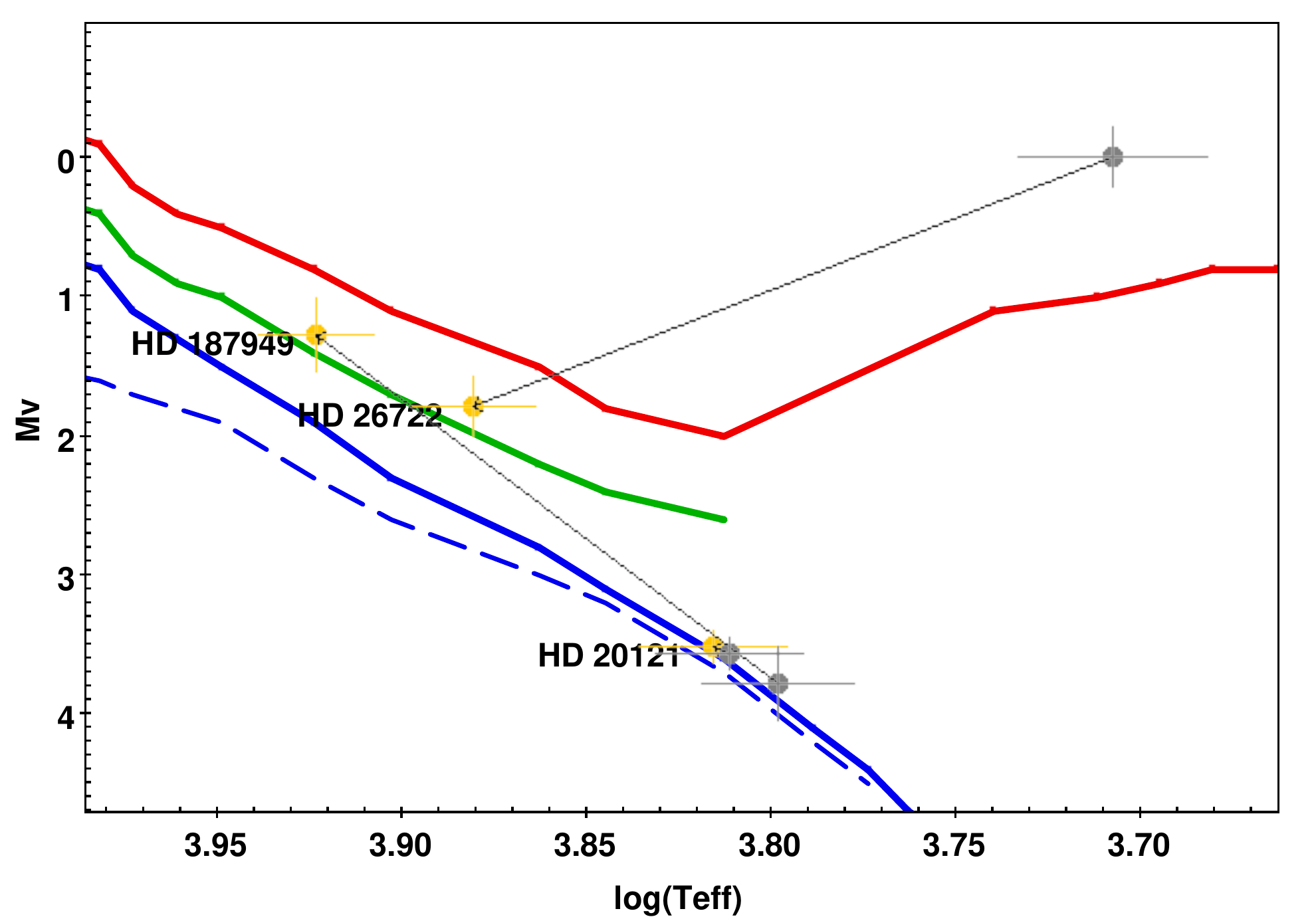}
    \caption{HD 20121, HD 26722, and HD 187949 systems on the HRD.
Solid curves represent main sequence, subgiant and
giant sequences (blue, green and red, respectively),
and dashed blue curve represents ZAMS~\citep{1992msp..book.....S}.
Yellow and grey circles represent primary (hotter)
and secondary components of the binaries, respectively, with uncertainty bars.
Left and right panels: Coelho and Phoenix stellar models, respectively.}
    \label{fig:HRD1}
\end{figure}

\begin{figure}[]
    \includegraphics[width=8cm]{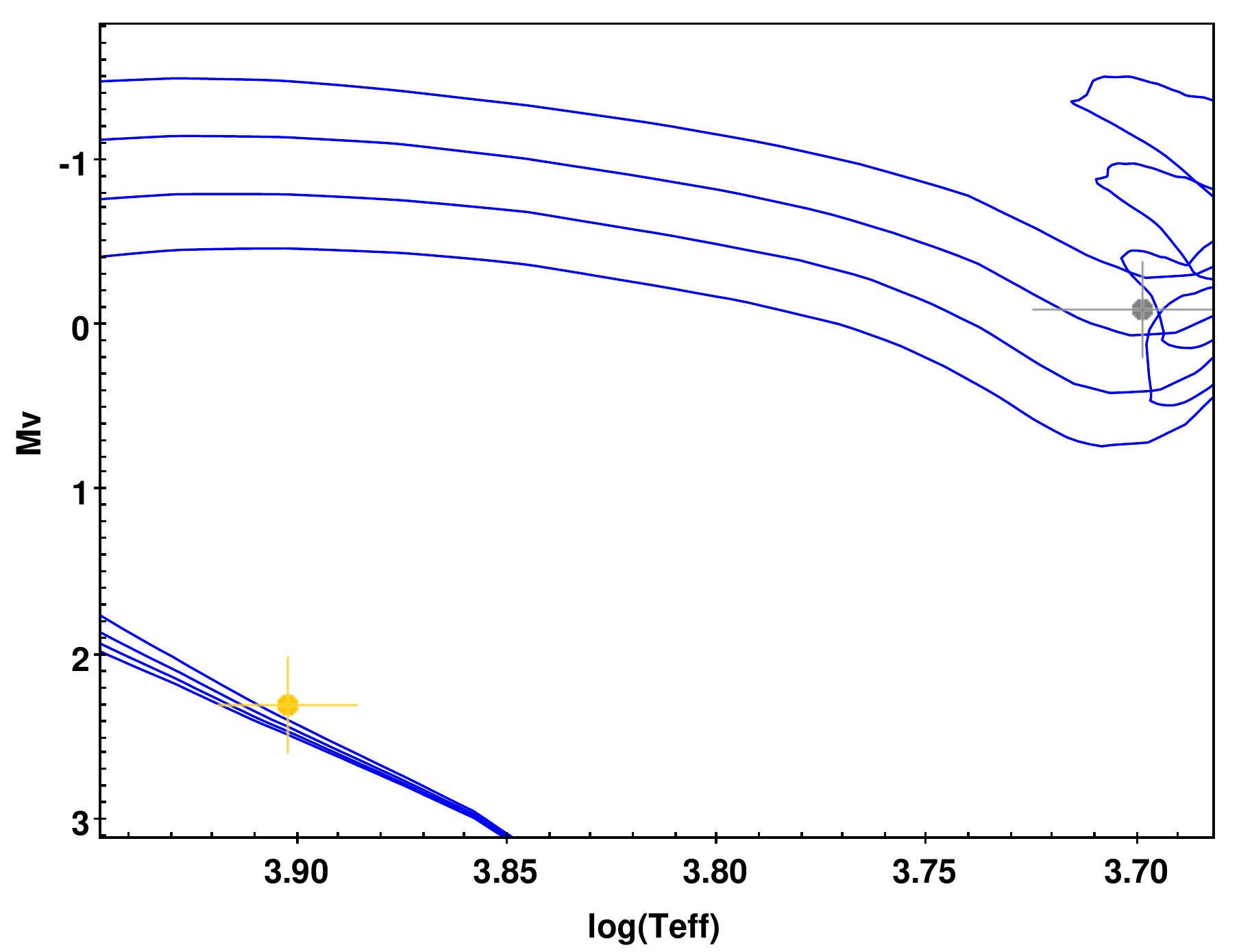}
    \caption{HD 26722 system (solution for Coelho stellar model) on the HRD.
        Thin blue curves are
        isochrones from~\cite{2012A&A...537A.146E} for ages (from top to bottom)
        $\log \tau=8.3,8.4,8.5,8.6$ yr, ``no rotation'' case, initial metallicity Z=0.014.
        Other designations are the same as in Fig.~\ref{fig:HRD1}.}
    \label{fig:HRD2}
\end{figure}

For the analysis of reduced data we relied on the software package Fitting Binary Stars 
\citep[{\tt fbs};][]{2020RAA....20..119K}.
This software was developed for the analysis of stellar spectra of binary systems
and was used previously by us in different studies
\citep{2019AstBu..74..183B,2019MNRAS.482.4408G,2020Ap&SS.365..169K,2021MNRAS.503.3856G,2022OAst...31..327M}.

This package was designed to fit observed stellar spectra with a library of theoretical stellar spectra
to determine velocities and values of stellar parameters like
($T_\mathrm{eff}$, $\log g$, $\mathrm{v \sin i}$, [Fe/H], W)$_{1,2}$
for both components in the binary system,
where parameter W$_{1,2}$ is the weight of each spectrum ($\rm W_1 + W_2 = 1$) in $V$-band (at the wavelength $\lambda = 5550$~\AA).
Additionally, the $E(B-V)$ value could be found in case spectra were corrected for the sensitivity of the observational system.
Different theoretical stellar libraries could be used with the {\tt fbs} software,
but we utilized usual stellar models from \citet{Coelho14,phoenix,tlusty}.

As was noted previously by \citet{2020Ap&SS.365..169K},  {\tt fbs} software can underestimate the real errors.
Since \'echelle spectra consist of hundreds thousands of points, the minimization of the function described in \citet{2020RAA....20..119K} is a very time-consuming process and for that reason 
Monte-Carlo simulations or use of Markov chain Monte Carlo methods are not implemented.
\citet{2020RAA....20..119K} suggested executing {\tt fbs} for each \'echelle spectrum separately
\citep{2020Ap&SS.365..169K,2022OAst...31..327M}.
For that reason we use {\tt fbs} in the same way here and our results are presented in Table~\ref{tab:stars} and displayed in Figure~\ref{fig:spectrum}. 
We analyzed all spectra with {\tt fbs} independently, 
and present parameters and their errors, which are average values for each star.
We produced this analysis with both \citet{Coelho14} and \citep{phoenix} stellar model.
Both libraries were convolved to match the HRS MR instrumental resolution.
These results are also shown in Table~\ref{tab:stars}.
Finally, after comparing these results, we use errors for each parameter in this work;
they are expressed in the last row of Table~\ref{tab:stars}.

\section{Parameters of the studied systems and discussion}
\label{txt:results}

The parameters of four systems observed with SALT are displayed in Figures~\ref{fig:HRD1} and~\ref{fig:HRD2} and are presented in Tables~\ref{tab:pairs} and~\ref{tab:stars}.
Absolute magnitudes of the components $M_V$ are calculated from parallax and visual brightness
(see Table~\ref{tab:pairs}) and from weight in $V$ band and interstellar reddening $E(B-V)$ values
(see Table~\ref{tab:stars}).

We were unable to detect a secondary component in the spectrum of HD\,114330,
which is why this system is not shown in the Figures.
One can see from Fig.~\ref{fig:HRD1} that both components of
HD\,20121 and HD\,187949 belong to the main sequence, within observational errors,
or are not far away from it.
Note that the catalogued spectral types of the HD\,187949 components (A2V+F7V)
fit the Coelho solution slightly better than the Phoenix solution.

On the contrary, the more luminous component of HD\,26722 looks more evolved than the hotter one.
However, as can be seen from Fig.~\ref{fig:HRD2}, both components fit
the isochrones~\cite{2012A&A...537A.146E}. We estimate the age of the system to be
$\log \tau=8.4-8.6$~yr, and masses of the components are
$m_g$~=~3.18$\pm$0.16~$m_\odot$ (the cooler component is a more luminous giant)
and $m_{ms}$~=~1.70$\pm$0.02~$m_\odot$ (the hotter component is a less luminous MS-star).

This preliminary list was compiled from the ORB6 catalog.
We have finished the inspection of the southern sky and,
given our previous research, where we found one candidate for systems
with non-coeval components, we can roughly estimate the proportion
of systems suspected of being formed by capture to be no more than 0.06 per cent.
We now plan to continue our research on the northern sky.

Note that another indicator that a binary system has formed as a result
of capture could be the observed difference in the chemical abundance
of the components. Such an effect was found for the RR Lyn system~\citep{2001ARep...45..888K}
and probably for V966 Per (Volkov 2022, private communication).
In the latter case, the components,
being placed on the Hertzsprung-Russell diagram (HRD) and evolutionary tracks, in addition
exhibit very different ages. However, both systems are relatively close
(period P=9.945 days and P=4.309 days, respectively) eclipsing binaries,
so it is possible that the system exchanged masses in the past.

\section{Conclusion}
\label{txt:summary}

We have spectroscopically studied the last four stars in the southern sky 
from our preliminary list of candidates for wide pairs with non-coeval components, and we have
found no evidence of non-coevality. The study will be continued on the northern sky.


\begin{acknowledgements}
We thank Andrew Tokovinin for his comments on the nature of HD~187949~A
and Igor Volkov for the information on his preliminary conclusions on V966~Per.
We also acknowledge Marina Ushakova and Vladimir Goradzhanov for their help
in finding and processing the data for HD~187949.
All spectral observations reported in this paper were obtained with
the Southern African Large Telescope (SALT) under program 2020-1-MLT-002 (PI: Alexei Kniazev) and
support from the National Research Foundation (NRF) of South Africa. 
This research was supported by the Ministry of Science and Higher Education of the Russian Federation grant 075-15-2022-262 (13.MNPMU.21.0003).
This research has made use of NASA's Astrophysics Data System,
of the SIMBAD database, operated at CDS (Strasbourg, France),
of TOPCAT, an interactive graphical viewer and editor for tabular data \citep{2005ASPC..347...29T}.
The acknowledgements were compiled using the Astronomy Acknowledgement Generator. 
\end{acknowledgements}

\bibliographystyle{raa}
\bibliography{noncoev}

\label{lastpage}
\end{document}